\documentstyle[12pt]{article}
\catcode`\@=11

\@addtoreset{equation}{section}
\@addtoreset{equation}{subsection}
\def\theequation{\ifnum\value{subsection}>0\relax
\thesubsection.\arabic{equation}\relax
\else\ifnum\value{section}>0\relax
\thesection.\arabic{equation}\relax
\else\arabic{equation}\fi\fi}

\newcommand{\ba}{\begin{eqnarray}}   %   these lines just define
\newcommand{\ea}{\end{eqnarray}}     %   equation and eqnarray commands
\newcommand{\be }{\begin{equation}}   %   insert them in text if you already
\newcommand{\ee }{\end{equation}}     %   have \be, \ba etc.

\newcommand{\RMA}{{\cal R}}     % universal R-matrix
\newcommand{\al}{\alpha}
\newcommand{\bet}{\beta}

\newcommand{\Ga}{\Gamma}

\newcommand{\Dl}{\Delta}
\newcommand{\si}{\sigma}

\newcommand{\lam}{\lambda}

\newcommand{\veps}{\varepsilon}

\begin{document}
%\begin{titlepage}
\rightline{  }
\vskip 1cm
\centerline{\Large \bf Boundary $K$-matrices  and the Lax pair }
\vskip 0.8cm
\centerline{\Large \bf for  1D open $XYZ$ spin-chain}

\vskip 0.9cm
\centerline{\bf Guo-xing JU$^{1,3}$, Shi-kun WANG$^2$, Ke WU$^1$, Chi XIONG$^{1}$  
}

\vskip0.5cm
\centerline{\sf 
1.Institute of Theoretical Physics}
\centerline{\sf 
Academia Sinica}
\centerline{\sf 
Beijing 100080,China}          
\centerline{\sf 
2.CCAST(World Laboratory),Beijing 100080,China \\
and \\
Institute of Applied Mathematics}
\centerline{\sf  
Academia Sinica}
\centerline{\sf  
Beijing 100080,China}                               
\centerline{\sf   
3.Physics Department}
\centerline{\sf  
Henan Normal University}
\centerline{\sf  
Xinxiang,Henan Province 453002,China}                                                       
\vskip0.5cm
\begin{abstract}
We analysis the symmetries of the reflection equation for  open $XYZ$ model 
and find  their solutions $K^{\pm}$ case by case. In the general open boundary
conditions, the Lax pair for  open one-dimensional $XYZ$ spin-chain is given.

\end{abstract}
\vskip0.5cm
%\end{titlepage}

\newpage

\section{Introduction}
In the formalism of the quantum inverse scattering method
(QISM), the integrability of a system with periodic boundary conditions is 
based on  the Yang-Baxter equation\cite{fad,kor}. Of central importance in 
the QISM is the ${\RMA}$-matrix. Relying on the previous  work of Cherednik
\cite{che}, Sklyanin generalized the QISM to system on a finite internal 
with independent boundary conditions on each end\cite{skl}. For such a system, 
the boundary conditions are  defined 
by the operators  $K^{\pm}$, which satisfy the so-called reflection 
equations (or called the boundary Yang-Baxter equations). The integrability conditions for systems with open boundary conditions 
are that both the Yang-Baxter equations and the reflection equations must be 
satisfied. 
Since Sklyanin's work,  a lot of integrable models with open boundaries, 
both in quantum field theory 
and in statistical mechanics, as the Toda chain\cite{skl}, nonlinear Schr\"odinger
 equation\cite{skl}, the $A_2^{(2)}$ spin chain with $U_q [SU(2)]$ 
symmetry\cite{mez}, supersymmetric $t-J$  model\cite{foe,gon}, sine-Gordon 
model\cite{gho}, $A_{n-1}$ vertex  models\cite{vega1}, the $XYZ$ 
model\cite{skl,vega2,ina}, have been studied extensively.\\

In this paper, we will present the solutions $K^{\pm}(u)$ to the reflection 
equations of open $XYZ$ model. The solutions $K^{\pm}(u)$ for $XYZ$ 
model have been 
studied by de Vega, Gonzalez-Ruiz and Inami\cite{vega2,ina}. However, 
we will persue a new procedure to the problem. Firstly, we analysis 
the  symmetries of the component forms of the reflection equations for $XYZ$ 
model and determine the equations to be solved. Then we study the chosen 
equations  to get the solutions $K^{\pm}$ case by case and their corresponding 
boundary terms in the Hamiltonians. For the case of $tr K^{+}(0)=0$, we get a 
new $K(u)$ that is not given in ref.\cite{vega2}(see (2.18) and (2.25)). Furthermore, we find 
that several $K$-matrices may correspond to the same boundary terms if they 
differ by a scaling function satisfied some conditions. We expect our procedure
and discussions to be helpful
to the understanding of the role played by $K$-matrices in the description
of a system. In the formalism
of QISM, the systems described by the above Hamiltonians with boundary terms
are integrable.\\
 
However , in the case of  periodic boundary condition, there is  an equivalent 
approach, which is called  the Lax representation \cite{kor}, to the proof of 
the integrability of the system. In this formalism, a model is said to be
completely
integrable if we can find a Lax pair such that the Lax equation is equivalent 
to the equation of motion of the model. Now, it is natural that one expect a 
variant of the Lax representation to systems with open boundary conditions. In 
fact, this has been
done, for example, for the one-dimensional (1D) Heisenberg open $XXZ$  chain
\cite{zhou}, the Hubbard model\cite{guan}. We will construct the Lax pair for 
the 1D open $XYZ$ spin-chain.\\

The paper is organized as follows. In section 2, after briefly reviewing the 
${\RMA}$-matrix for the $XYZ$ model, the corresponding Yang-Baxter 
equation and the reflection equations, we introduce the 
symmetries for the reflection equations and find the $K$-matrices case by case.
In the second half of this section, we give the corresponding boundary terms
for each solution $K^{\pm}$ and discuss a symmetry of the general Hamiltonian
for the open $XYZ$ model.
In section 3, we will construct the Lax pair for the $XYZ$ model with 
general boundary conditions. We will make some discussions in the last section.\\

\section{Solutions to the reflection equations for the $XYZ$ model}
The $XYZ$ model is defined in terms of the Boltzmann weights given by the elliptic 
eight-vertex solution of the Yang-Baxter equation:
\be
{\RMA}_{12}(u){\RMA}_{13}(u+v){\RMA}_{23}(v)={\RMA}_{23}(v){\RMA}_{13}(u+v){\RMA}_{
12}(u)
\ee
Here ${\RMA}_{12}(u)={\RMA}(u)\otimes 1, {\RMA}_{23}(u)=1\otimes {\RMA}(u)$ . The 
${\RMA}$-matrix ${\RMA}(u)$ is given as follows:
\be
{\RMA}(u)=\left (
\begin{array}{cccc}
{\rm sn} (u+\eta)   &0      &0      & k{\rm sn}\eta {\rm sn} u {\rm sn} (u+\eta)\\
0            &{\rm sn} u &{\rm sn} \eta &0\\
0                  &{\rm sn} \eta &  {\rm sn} u &0\\
k{\rm sn}\eta {\rm sn} u {\rm sn} (u+\eta)&0 &0 & {\rm sn} (u+\eta) 
\end{array}
\right )
\ee
where {\rm sn} (and {\rm cn},{\rm dn} in the formulas below) is the Jacobi elliptic 
function of modulus $0\leq k\leq 1$. For the properties of the ${\RMA}$-matrix 
(2.2) 
see refs.\cite{vega2,ina}.\\

To find boundary terms compatible with integrability, Sklyanin has introduced a 
pair of boundary $K$-matrices $K^{\pm}$\cite{skl}. $K^{\pm}$ are the solutions of 
the 
following reflection equations:
\be
{\RMA}_{12}(u-v)K_1^{-}(u){\RMA}_{12}(u+v)K_2^{-}(v)=
K_2^{-}(v){\RMA}_{12}(u+v)K_1^{-}(u){\RMA}_{12}(u-v)
\ee
\be
{\RMA}_{12}(-u+v)K_1^{+t_1}(u){\RMA}_{12}(-u-v-2\eta)K_2^{+t_2}(v)=
K_2^{+t_2}(v){\RMA}_{12}(-u-v-2\eta)K_1^{+t_1}(u){\RMA}_{12}(-u+v)
\ee
where $K_1^{\pm}=K^{\pm}\otimes 1, K_2^{\pm}=1\otimes K^{\pm}$. Here we should note
that equations (2.3) and (2.4) only hold for the symmetric ${\RMA}$-matrix(i.e. 
${\RMA}$ has 
both $P$ and $T$ symmetry). The ${\RMA}$-matrix given in equation (2.2) is 
symmetric.\\

For a solution $K^{-}$ of the equation (2.3), then
\be
K^{+}(u)=K^{-t}(-u-\eta)
\ee
gives the solution to the equation (2.4). So it is sufficient to consider the 
equations (2.3). Let $K^{-}(u)$ be of the following form:
\be
K^{-}(u)=\left(
\begin{array}{cc}
a_1(u) &a_2(u)\\
a_3(u) &a_4(u)
\end{array}
\right)
\ee
For the ${\RMA}$-matrix (2.2), the equation (2.3) in component forms is 
equivalent to the following 12 equations:
\be
\left.
\begin{array}{rcl}
    d_1 e_2 u_2 v_2 + c_1 u_3 v_2 - c_1 u_2 v_3 - d_1 e_2 u_3 v_3 &=&  0,\\
    d_2 e_1 u_2 v_2 + c_2 u_3 v_2 - c_2 u_2 v_3 - d_2 e_1 u_3 v_3 &=&  0,
\end{array}
\right\}
\ee

\be
\left.
\begin{array}{rcl}
  e_2 u_1 v_1 + e_1 u_4 v_1 - e_1 u_1 v_4 - e_2 u_4 v_4 &=&  0,\\    
  c_1 d_2 u_1 v_1 + c_2 d_1 u_4 v_1 - 
      c_2 d_1 u_1 v_4 - c_1 d_2 u_4 v_4 &=&  0, 
\end{array}
\right\}
\ee 

\be
\left.
\begin{array}{rcl}   
 c_2 e_1 u_2 v_1 - c_1 e_2 u_2 v_1 - d_1 u_3 v_1  + 
      d_2 u_3 v_1  - e_1 u_1 v_2 - e_2 u_4 v_2  + c_2 d_1 u_1 v_3  + 
      c_1 d_2 u_4 v_3  &=&  0,\\
 c_2 e_1 u_2 v_4 - c_1 e_2 u_2 v_4 - d_1 u_3 v_4 +
      d_2 u_3 v_4  - e_1 u_4 v_2 - e_2 u_1 v_2  + c_2 d_1 u_4 v_3  +
      c_1 d_2 u_1 v_3  &=&  0,
\end{array}
\right\}
\ee

\be
\left.
\begin{array}{rcl}   
 c_2 e_1 u_3 v_1 - c_1 e_2 u_3 v_1 - d_1 u_2 v_1 + d_2 u_2 v_1
     - e_1 u_1 v_3  - e_2 u_4 v_3 + c_2 d_1 u_1 v_2 + 
      c_1 d_2 u_4 v_2  &=&  0,\\
 
 c_2 e_1 u_3 v_4 - c_1 e_2 u_3 v_4 - d_1 u_2 v_4 + d_2 u_2 v_4 
     - e_1 u_4 v_3  - e_2 u_1 v_3 + c_2  d_1 u_4 v_2 +
      c_1 d_2 u_1 v_2  &=&  0,
\end{array}
\right\}
\ee

\be
\left.
\begin{array}{rcl} 
 d_2 e_1 u_2 v_1 + c_2 u_3 v_1 - u_1 v_3  + d_1 d_2 u_1 v_3 + 
      c_1 c_2 u_4 v_3 - e_1 e_2 u_4 v_3 - 
      d_1 e_2 u_2 v_4 - c_1 u_3 v_4 &=&  0,\\
 d_2 e_1 u_2 v_4 + c_2 u_3 v_4 - u_4 v_3  + d_1 d_2 u_4 v_3 +
      c_1 c_2 u_1 v_3 - e_1 e_2 u_1 v_3  -
      d_1 e_2 u_2 v_1 - c_1 u_3 v_1  &=&  0,
\end{array}
\right\}
\ee

\be
\left.
\begin{array}{rcl}
  d_2 e_1 u_3 v_1 + c_2 u_2 v_1 - u_1 v_2 + 
      d_1 d_2 u_1 v_2 +  c_1 c_2 u_4 v_2 - 
      e_1 e_2 u_4 v_2  - d_1 e_2 u_3 v_4  - c_1 u_2 v_4  &=&  0,\\
  d_2 e_1 u_3 v_4 + c_2 u_2 v_4 - u_4 v_2 +
      d_1 d_2 u_4 v_2 +  c_1 c_2 u_1 v_2 -
      e_1 e_2 u_1 v_2  - d_1 e_2 u_3 v_1  - c_1 u_2 v_1  &=&  0,
\end{array}
\right\}
\ee
where $u_i=a_i(u), v_i=a_i(v) (i=1,2,3,4)$ and
\be
\left.
\begin{array}{ll}
c_1=\frac{{\rm sn}\eta}{{\rm sn}(u+v+\eta)},  &c_2=\frac{{\rm sn}\eta}{{\rm sn}(u-v+\eta)},\\
d_1=\frac{{\rm sn}(u+v)}{{\rm sn}(u+v+\eta)}, &d_2=\frac{{\rm sn}(u-v)}{{\rm sn}(u-v+\eta)},\\
e_1=k {\rm sn}\eta {\rm sn}(u+v),&   e_2=k {\rm sn}\eta {\rm sn}(u-v).
\end{array}
\right\}
\ee
After analyzing the system of the equations (2.7) -(2.12), we find that there are
three kinds of symmetries for the solutions of equation (2.3).\\
(A){\it The Scaling Symmetry.} Multiplication of the solution $K^{-}(u)$ by an 
arbitrary function $f(u)$ is still a solution of equation (2.3).\\
(B){\it Symmetry of Spectral Parameter.} If we take a new spectral parameter 
$\bar{u}= \mu u$ where $\mu$ is a complex constant, the new matrix $K^{-}(\bar{u})$
is still a solution of (2.3). This symmetry is useful when we consider the 
rational limit of the matrix $K^{-}(u)$.\\
(C){\it Symmetry of Interchanging Variables u and v with Different Indices}. Under the exchange
of $u_2\leftrightarrow u_3$ and $v_2\leftrightarrow v_3$, the equations
(2.7)  are invariant ,while the equations (2.9), (2.11) are turned into the equations (2.10) and
(2.12), respectively. On the other hand, upon the exchange of $u_1\leftrightarrow u_4$ and 
$v_1\leftrightarrow v_4$, the equations (2.8) are invariant, while for each set of equations 
(2.9)-(2.12), one equation is changed into another.\\
 
 In view of the relation among $c_i, d_i, e_i (i=1,2)$
 \be
\frac{d_1 e_2}{c_1}=\frac{d_2 e_1}{c_2},
\ee
two equations in (2.7) and (2.8) are equivalent, respectively. Because of this fact and
the above symmetries, we only need choose four equations from equations (2.7)-(2.12) in order
to solve the reflection equation (2.3). In the following, we will take the first equation of
(2.7), (2.8), (2.9) and (2.11), respectively. Of the four chosen equations, three equations are
necessary to determine the $K^{-}$ due to the symmetry (A). The fourth equation may be as the
consistency condition for the solutions of the three equations.\\

The solutions can be classified according to whether $a_1$ and $a_2$ equal to zero or not. If 
assuming $a_2\not=0$, we differentiate both sides of the first equation of (2.7) with 
respect to the spectral  variable $v$ and set $u=v$, then we get the following equation
\be
\frac{d}{du}(\frac{u_3}{u_2})=-k{\rm sn}(2u)[1-(\frac{u_3}{u_2})^2].
\ee
Similarly, assuming $a_1\not=0$,one also has
\be
\frac{d}{du}(\frac{u_4}{u_1})=-\frac{1}{{\rm sn}(2u)}[1-(\frac{u_4}{u_1})^2].
\ee
if we differentiate both sides of the first equation of (2.8) with respect to $v$ and 
set $u=v$. Now we solve the four equations chosen above case by case. The procedure we 
adopt below is the same as that in refs.\cite{vega2,ina}. We give the main results in the 
following.\\

{\it Case (a): $a_1=0$}. Using the first equation of (2.8), we have $a_4=0$. The nontrivial cases 
correspond to $a_2\not=0$ and $a_3\not=0$ which come from the first equation of (2.7). It has 
the following two subcases.\\
{\it Subcase (a1): $(\frac{u_3}{u_2})^2=1$}. From equation (2.15) and the symmetry (A), we obtain
\be
K^{-}(u)=\left(
\begin{array}{cc}
0 &1\\
\pm 1 &0
\end{array}
\right)
\ee
{\it Subcase (a2): $(\frac{u_3}{u_2})^2\not=1$}. Integrating equation (2.15) and taking account of 
the symmetry (A), we have
\be
K^{-}(u)=\left(
\begin{array}{cc}
0          &\lambda_{-}(1-k{\rm sn}^2 u)+1+k {\rm sn}^2 u\\
\lambda_{-}(1-k{\rm sn}^2 u)-1-k {\rm sn}^2 u    &0
\end{array}
\right)
\ee
here $\lambda_{-}$ is an arbitrary constant.\\

{\it Case (b): $a_2=0$}. From the first equation of (2.7), we have $a_3=0$, so the nontrivial 
cases are those of $a_1\not=0$ and $a_4\not=0$ from the first equation of (2.8).Here we also have 
two subcases.\\
{\it Subcase (b1): $(\frac{u_4}{u_1})^2=1$}. From equation (2.16) and the symmetry (A), we get
\be
K^{-}(u)=\left(
\begin{array}{cc}
1 &0\\
0 &\pm 1 
\end{array}
\right)
\ee
{\it Subcase (b2): $(\frac{u_4}{u_1})^2\not=1$}. Considering the symmetry (A) and integrating 
equation (2.16), we have
\be
K^{-}(u)=\frac{1}{{\rm sn}\xi_{-}}
\left(
\begin{array}{cc}
{\rm sn}(\xi_{-}+u) &0\\
0 &{\rm sn}(\xi_{-}-u) 
\end{array}
\right)
\ee
here $\xi_{-}$ is an arbitrary constant.\\

{\it Case (c): $a_1\not=0, a_2\not=0$}.Using the first equation of (2.7) and (2.8), 
respectively, we know that $a_3\not=0, a_4\not=0$. For this case, we may have following
four subcases.\\
{\it Subcase (c1): $(\frac{u_3}{u_2})^2=1, (\frac{u_4}{u_1})^2=1$}, i.e. $u_3=\varepsilon_2 u_2,
u_4=\varepsilon _1 u_1, \varepsilon _1^2=\varepsilon _2^2=1$. Using the first equation of (2.9), 
we get two $K^{-}$ solutions:
\be
K^{-}_A(u)=
\left(
\begin{array}{cc}
1+\varepsilon_2 k{\rm sn}^2 u &\al_{-}{\rm sn} u\\
\varepsilon_2\al_{-}{\rm sn} u &1+\varepsilon_2 k{\rm sn}^2 u  
\end{array}
\right)\;\;\;(for\;\;\varepsilon_1=1)
\ee
\be
K^{-}_B(u)=
\left(
\begin{array}{cc}
1+\varepsilon_2 k{\rm sn}^2 u &\bet_{-}{\rm cn} u{\rm dn} u\\
\varepsilon_2\bet_{-}{\rm cn} u {\rm dn} u&-1-\varepsilon_2 k{\rm sn}^2 u  
\end{array}
\right)\;\;\;(for \;\;\varepsilon_1=-1)
\ee
where $\al_{-}, \bet_{-}$ are two arbitrary constants.\\
{\it Subcase (c2): $(\frac{u_3}{u_2})^2=1, (\frac{u_4}{u_1})^2\not=1$},Then $u_3=\varepsilon_2 u_2,
\varepsilon _2^2=1$. Solving the first equation of (2.9), we have the following solution,
\be
K^{-}(u)=\frac{1}{{\rm sn}\xi_{-}}
\left(
\begin{array}{cc}
{\rm sn}(\xi_{-}+u)(1+\varepsilon_2 k{\rm sn}^2 u)    &
\frac{\al_{-} {\rm sn}u {\rm cn}u {\rm dn} u}{1-k^2{\rm sn}^2\xi_{-}{\rm sn}^2 u}\\
\frac{\varepsilon_2 \al_{-} {\rm sn}u {\rm cn}u {\rm dn} u}{1-k^2{\rm sn}^2\xi_{-}{\rm sn}^2 u}
 &{\rm sn}(\xi_{-}-u)(1+\varepsilon_2 k{\rm sn}^2 u )
\end{array}
\right)
\ee
here $\xi_{-}$ and $\al_{-}$ are two arbitrary constants.\\
{\it Subcase (c3): $(\frac{u_3}{u_2})^2\not=1, (\frac{u_4}{u_1})^2=1$}. Taking $u_4=
\varepsilon_1 u_1$ and solving the first equation of (2.9), we get two $K^{-}$ solutions,
\ba
K^{-}_A(u)&=&
\left(
\begin{array}{cc}
{\rm cn}u{\rm dn}u &\mu_{-} {\rm sn}(2u)[(1-k{\rm sn}^2 u)\lambda_{-}+1+k{\rm sn}^2 u]\\
\mu_{-} {\rm sn}(2u)[(1-k{\rm sn}^2 u)\lambda_{-}-1-k{\rm sn}^2 u]& {\rm cn}u{\rm dn}u
\end{array}
\right)\nonumber\\
&&(for \;\;\varepsilon_1=1)
\ea

\ba
K^{-}_B(u)&=&
\left(
\begin{array}{cc}
{\rm sn}u &\mu_{-} {\rm sn}(2u)[(1-k{\rm sn}^2 u)\lambda_{-}+1+k{\rm sn}^2 u]\\
\mu_{-} {\rm sn}(2u)[(1-k{\rm sn}^2 u)\lambda_{-}-1-k{\rm sn}^2 u]& -{\rm sn}u
\end{array}
\right)\nonumber\\
&&(for \;\;\varepsilon_1=-1)
\ea
here $\mu_{-}, \lambda_{-}$ are two arbitrary constants.\\
{\it Subcase (c4): $(\frac{u_3}{u_2})^2\not=1, (\frac{u_4}{u_1})^2\not=1$}.This 
is the most general case. Now we first have to integrate equations (2.15) 
and (2.16), then insert the results into the first
equation of (2.9). Upon the use of the symmetry (A), we find that $K^{-}$ has 
the following form:
\be
K^{-}(u)=\frac{1}{{\rm sn}\xi_{-}}
\left(
\begin{array}{cc}
{\rm sn}(\xi_{-}+u) &
\mu_{-} {\rm sn}(2u)\frac{(1-k{\rm sn}^2 u)\lambda_{-}+1+k{\rm sn}^2 u}{1-k^2{\rm sn}^2\xi_{-}{\rm sn}^2 u}\\
\mu_{-} {\rm sn}(2u)\frac{(1-k{\rm sn}^2 u)\lambda_{-}-1-k{\rm sn}^2 u}{1-k^2{\rm sn}^2\xi_{-}{\rm sn}^2 u}
&{\rm sn}(\xi_{-}-u) 
\end{array}
\right)
\ee
here $\xi_{-}, \mu_{-}, \lambda_{-}$ are arbitrary constants. \\

Note that the subcases (a1),(b1) and (c1) are discussed in ref.\cite{vega2},
while the subcase (c4) is studied in ref.\cite{ina}.\\

Now we consider the boundary terms corresponding to the $K$-matrices given in equations
(2.17)-(2.26). For the case of $tr K^{+}(0)\not=0$, the Hamiltonians with boundary 
terms are obtained from the first derivative of the transfer matrix
\cite{skl,vega2,ina}
\be
H=2r(\eta)(\sum_{n=1}^{N-1}H_{n,n+1}+\frac{1}{2}K_1^{-}(0)^{-1}K_1^{-\prime}(0)
+\frac{tr_0K_0^{+}(0)H_{N0} }{trK^{+}(0)})
\ee
where $K^{-}(0)^{-1}$ in the second term is introduced for the case when 
$K^{-}(0)\not=1$,
$r(\eta)={\rm sn}\eta$ and the two-site Hamiltonian $H_{n,n+1} $ is given by
\be
H_{n,n+1}=\frac{1}{r(\eta)}P_{n,n+1}R^{\prime}_{n,n+1}(0)
\ee
$P_{i,j}$ is the permutation operator acting on $V_i\otimes V_j$ with $V_i=V_j\cong
V$, i.e. $P(x\otimes y)=y\otimes x \;\;(x,y\in V)$. If
\be
tr K^{+}(0)=0,
\ee
then as pointed out in ref.\cite{vega2}, we will not have a well defined Hamiltonian
(2.27) from the first derivative of the transfer matrix. In this case, we may obtain
a Hamiltonian from the second derivative of the transfer matrix. Only when the
condition
\be
tr_{0}[K_0^{+}(0) H_{N0}(0)]\propto 1
\ee
holds for the $K$-matrix, will such a Hamiltonian be that with nearest-neighbor 
interactions. Otherwise, it will contain terms that couple
every pair of sites in the bulk with the boundary, i.e. the Hamiltonian is non-local
\cite{vega2,cue}. The condition (2.29) holds for the $K$-matrices given in equations (2.17), (2.18), 
(2.19), (2.22) and (2.25), but the condition (2.30) does not hold for these $K$-matrices.
Due to the above reason, therefore, we will not 
discuss the Hamiltonians corresponding to these $K$-matrices below.\\

Now we focus on the case of $trK^{+}(0)\not=0$. Using the equations (2.2) and 
(2.8), the first term in the equation (2.27), which describes the bulk properties 
of the $XYZ$ model, is given as follows,
\be
H_{bulk}=\sum_{n=1}^{N-1}[(1+\Ga)\si^x_n\si^x_{n+1}+(1-\Ga)\si^y_n\si^y_{n+1}
+\Dl\si^z_n\si^z_{n+1}],
\ee
where
\be
\Ga=k {\rm sn}^2 \eta, \;\;\; \Dl={\rm cn}\eta {\rm dn}\eta.
\ee
and $\si_{n}^{x},\si_{n}^{y},\si_{n}^{z}$ are the Pauli matrices. The second 
and the third terms in the equations (2.27) represent the left and 
right boundary terms, respectively. Using the equations (2.2) and (2.28), the 
boundary terms for the $K^{-}$-matrices (2.20), (2.21), (2.23), (2.24), (2.26)
and their corresponding $K^{+}$-matrices are of the following form,
\be
H_{boundary}={\rm sn}\eta (A_{-}\si_1^z + B_{-}\si_1^{+} +C_{-}\si_1^{-}
+A_{+}\si_N^z + B_{+}\si_N^{+} +C_{+}\si_N^{-})
\ee
with coefficients $A_{\pm}, B_{\pm},C_{\pm}$ being as follows, respectively,
\be
\left.
\begin{array}{r}
A_{\pm}=\frac{{\rm cn}\xi_{\pm} {\rm dn}\xi_{\pm}}{{\rm sn}\xi_{\pm}},\\
B_{\pm}=C_{\pm}=0;
\end{array}
\right\}\;\;\;( for \;\;subcase\;(b2))
\ee
\be
\left.
\begin{array}{rr}
& A_{\pm}=0,\\
B_{+}=\veps_2\al_{+},&B_{-}=\al_{-},\\
C_{+}=\al_{+}, &C_{-}=\veps_2\al_{-};
\end{array}
\right\}\;\;\;( for \;\;subcase\;(c1))
\ee
\be
\left.
\begin{array}{rr}
& A_{\pm}=\frac{{\rm cn}\xi_{\pm} {\rm dn}\xi_{\pm}}{{\rm sn}\xi_{\pm}},\\
B_{+}=-\frac{\veps_2\al_{+}}{{\rm sn}\xi_{+}},& B_{-}=\frac{\al_{-}}{{\rm sn}\xi_{-}},\\
C_{+}=-\frac{\al_{+}}{{\rm sn}\xi_{+}}, & C_{-}=\frac{\veps_2\al_{-}}{{\rm sn}\xi_{-}};
\end{array}
\right\}\;\;\;( for \;\;subcase\;(c2))
\ee
\be
\left.
\begin{array}{rr}
&A_{\pm}=0,\\
B_{\pm}=\mp 2\mu_{\pm}(\lam_{\pm}+1),&C_{\pm}=\mp 2\mu_{\pm}(\lam_{\pm}-1);
\end{array}
\right\}\;\;\;( for \;\;subcase\;(c3))
\ee

\be
\left.
\begin{array}{rr}
& A_{\pm}=\frac{{\rm cn}\xi_{\pm} {\rm dn}\xi_{\pm}}{{\rm sn}\xi_{\pm}},\\
B_{\pm}=\frac{2\mu_{\pm}(\lam_{\pm}+1)}{{\rm sn}\xi_{\pm}},&
C_{\pm}=\frac{2\mu_{\pm}(\lam_{\pm}-1)}{{\rm sn}\xi_{\pm}};
\end{array}
\right\}\;\;\;( for \;\;subcase\;(c4))
\ee
It is easy to see that the equation (2.20) is the special case $\mu_{-}=0$ of the
equation (2.26). Comparing the equation (2.34) with the equation (2.38), we know 
that the off-diagonal elements of the $K$-matrix are responsible for the $B_{\pm},
C_{\pm}$ terms in the $H_{boundary}$.\\

From the symmetry (A) and the equation (2.27), we know that if the scaling factor
$f(u)$ satisfies the conditions: $f^{\prime}(0)=0$ and $f(0)=const.$, then $f(u)$
will not change the boundary terms. $f(u)={\rm cn}u{\rm dn}u/(1-k^2{\rm sn}^2\xi
{\rm sn}^2 u)$ is such an example. When we take $\xi_{-}={\cal K}$(${\cal K}$ is 
the half-period magnitude of the Jacobi elliptic function) in the equation (2.23),
then the equation (2.23) reduces to the equation (2.21) up to a factor 
${\rm cn}u{\rm dn}u/(1-k^2{\rm sn}^2 u)$; Similarly, the equation (2.26) reduces 
to the equation (2.24) up to a factor $1/(1-k^2{\rm sn}^2 u)$. These two factors 
satisfies the conditions for the scaling factor $f(u)$ above. The equations 
(2.21) and (2.23), therefore, correspond to the same boundary terms upon taking 
$\xi_{-} ={\cal K}, \xi_{+}=-{\cal K}$; the same holds for the equations (2.24)
and (2.26).\\

In closing this section, we make a remark on the symmetry of the Hamiltonian (2.27).
It follows from the equations (2.31) and (2.33) that the Hamiltonian is 
transposition invariant if we make any replacement given below,
\be
\begin{array}{ll}
(1)&B_{-}\longrightarrow C_{-},\;\;B_{+}\longrightarrow C_{+};
\end{array}
\ee
\be
\begin{array}{ll}
(2)   &  a)\si_n\longrightarrow\si_{N-n+1},\;\;{\rm in\; particular},\;\;
   \si_1\longrightarrow\si_N,\;\;\si_2\longrightarrow\si_{N-1};\\
   &  b)A_{-}\longrightarrow A_{+},\;\;B_{-}\longrightarrow C_{+},\;\;
C_{-}\longrightarrow B_{+}.   
\end{array}
\ee
In the next section, we will find that the second replacement is very useful 
for the construction of the Lax pair of the open $XYZ$ spin-chain.\\

\section{The Lax pair for the open $XYZ$ spin-chain}

In section 2, we have derived the Hamiltonian of the open $XYZ$ 
spin-$\frac{1}{2}$ chain. For the 1D system with $N$ sites, the 
general form of the the Hamiltonian  is given by the equations (2.27), (2.31) 
and (2.33)\cite{ina},
\ba
H &=&-\frac{1}{2}\sum_{n=1}^{N-1}(J_x\si _{n}^{x}\si _{n+1}^{x}+J_y\si _{n}^{y}
\si _{n+1}^{y}+J_z\si _{n}^{z}\si _{n+1}^{z})   \nonumber\\
& &+{\rm  {\rm  sn}} \eta (A_{-}\si_{1}^{z}+
B_{-}\si_{1}^{+}+C_{-}\si_{1}^{-}+A_{+}\si_{N}^{z}
+B_{+}\si_{N}^{+}
+C_{+}\si_{N}^{-}),
\ea
where
\be
J_x=-2(1+k {\rm sn}^2 \eta), J_y=-2(1-k {\rm sn}^2 \eta),J_z=-2{\rm cn} \eta {\rm dn} \eta ,
\ee
The equations of  motion for the system (3.1) are given as follows,
\ba
\frac{d}{dt} \si_{n}^x &=& -J_y\si_n^z(\si_{n+1}^y+\si_{n-1}^y)
                      +J_z\si_n^y(\si_{n+1}^z+\si_{n-1}^z),\\
\frac{d}{dt} \si_{n}^y &=& -J_z\si_n^x(\si_{n+1}^z+\si_{n-1}^z)
                      +J_x\si_n^z(\si_{n+1}^x+\si_{n-1}^x),\\
\frac{d}{dt} \si_{n}^z &=& -J_x\si_n^y(\si_{n+1}^x+\si_{n-1}^x)
                      +J_y\si_n^x(\si_{n+1}^y+\si_{n-1}^y),\\
                       & &(n=2,\cdots,N-1) \nonumber
\ea
\ba
\frac{d}{dt} \si_{1}^x &=& -J_y\si_1^z\si_2^y+J_z\si_1^y\si_2^z
                       +{\rm  sn} \eta(-2A_{-}\si_1^y+i(B_{-}-C_{-})\si_1^z),\\
\frac{d}{dt} \si_{1}^y &=& -J_z\si_1^x\si_2^z+J_x\si_1^z\si_2^x
                       +{\rm  sn} \eta(2A_{-}\si_1^x-(B_{-}+C_{-})\si_1^z),\\
\frac{d}{dt} \si_{1}^z &=& -J_x\si_1^y\si_2^x+J_y\si_1^x\si_2^y
                       -2i{\rm  sn} \eta(B_{-}\si_1^{+}-C_{-}\si_1^{-}),
\ea
\ba
\frac{d}{dt} \si_{N}^x &=& -J_y\si_{N-1}^y\si_N^z+J_z\si_{N-1}^z\si_N^y
                       +{\rm  sn} \eta(-2A_{+}\si_N^y+i(B_{+}-C_{+})\si_N^z),\\
\frac{d}{dt} \si_{N}^y &=& -J_z\si_{N-1}^z\si_N^x+J_x\si_{N-1}^x\si_N^z
                       +{\rm  sn} \eta(2A_{+}\si_N^x-(B_{+}+C_{+})\si_N^z),\\
\frac{d}{dt} \si_{N}^z &=& -J_x\si_{N-1}^x\si_N^y+J_y\si_{N-1}^y\si_N^x
                       -2i{\rm  sn} \eta(B_{+}\si_N^{+}-C_{+}\si_N^{-}).
\ea
In order to rewrite equations above in the Lax form, we consider the following 
operator version of an auxiliary linear problem,
\ba
\Psi_{n+1} &=& L_n(u)\Psi_n , \;\;\;(n=1,2,\cdots,N) \\
\frac{d}{dt}\Psi_n &=& M_n(u)\Psi_n , \;\;\;(n=2,\cdots,N) \\
\frac{d}{dt}\Psi_{N+1} &=& Q_N(u)\Psi_{N+1}, \\
\frac{d}{dt}\Psi_1 &=& Q_1(u)\Psi_1, 
\ea
where $u$ is the spectral parameter which does not depend on the time $t$,
$L_n(u), M_n(u), Q_1(u)$ and $Q_N(u)$ are called the Lax pair. $Q_1(u)$ and
$Q_N(u)$ are responsible for the boundary conditions. The consistency conditions for 
the equations (3.12)-(3.15) are  the following Lax equations:
\ba
\frac{d}{dt}L_n(u) &=& M_{n+1}(u)L_n(u)-L_n(u)M_n(u), \;\;\; (n=2,\cdots,N-1)\\ 
\frac{d}{dt}L_N(u) &=& Q_N(u)L_N(u)-L_N(u)M_N(u),\\
\frac{d}{dt}L_1(u) &=& M_2(u)L_1(u)-L_1(u)Q_1(u).
\ea
From the eqs. (3.3)-(3.5) and (3.16), we see that they are the same as those in the case of periodic 
boundary condition. For the $XYZ$ model with periodic boundary condition, the 
Lax pair $L_n, M_n$  has been given by Sogo and Wadati\cite{sogo}. The operators 
$L_n$ and $M_n$ are of the following forms,
\be
L_n=\left (
\begin{array}{cc}
w_4 + w_3\si_n^z          &   w_1\si_n^x -i w_2\si_n^y\\
w_1\si_n^x +i w_2\si_n^y&   w_4 - w_3\si_n^z
\end{array}
\right )
\ee
\be
M_n=\left (
\begin{array}{cc}
M_n^0+M_n^3   & M_n^1-i M_n^2\\
M_n^1+i M_n^2 & M_n^0-M_n^3
\end{array}
\right )
\ee
where
\be
\begin{array}{l}
M_n^0=\sum_{i=1}^{3}F_i\si_n^i\si_{n-1}^i,\\
M_n^k=G_k(\si_n^k+\si_{n-1}^k)+\sum_{i=1}^{3}\sum_{j=1}^{3}\epsilon ^{ijk}H_k\si_n^i\si_{n-1}^j,
\end{array}
\ee
\be
\begin{array}{ll}
w_1(u)+w_2(u)={\rm  sn} \eta,  &w_1(u)-w_2(u)=k {\rm  sn} \eta {\rm  sn} u {\rm  sn}(u+ \eta),\\
w_3(u)+w_4={\rm  sn}(u+ \eta), &w_4(u)-w_3={\rm  sn} u,
\end{array}
\ee
and $\si^0=1$, $\si^1, \si^2, \si^3=\si^x, \si^y, \si^z$ are the Pauli matrices. For 
the explicit expressions of $F_i, G_i,H_i$ see ref. \cite{sogo}. So, for the 
open $XYZ$ model, we only have to construct operators $Q_1(u)$ and $Q_N(u)$. 
If we write $Q_1(u)$ and $Q_N(u)$ in the following forms:
\be
Q_1(u)=(a_{ij}^{(-)})_{2\times 2},\;\;\; Q_N(u)=(a_{ij}^{(+)})_{2\times 2}
\ee
and
\be
a_{ij}^{(-)} = \sum_{l=0}^{3}a_{ij}^{(-)l}\si_1^l , \;\;\;\;
a_{ij}^{(+)} = \sum_{l=0}^{3}a_{ij}^{(+)l}\si_N^l ,
\ee
Substituting the equations (3.23) and (3.24) into the equations (3.17) and (3.18), we
obtain 32 equations for $a_{ij}^{(\pm)l} (i,j=1,2, l=0,1 , 2, 3)$. Due to the following 
lemma, we only have to solve 16 equations ,for example, for $a_{ij}^{(+)l} 
(i,j=1,2, l=0,1 , 2, 3)$. \\

{\it Lemma}: If we make the replacement (2.40), then we have
\be
a_{ij}^{(-)t}\longrightarrow a_{ij}^{(+)}
\ee
provided that the following correspondences hold
\be
\left.
\begin{array}{l}
(\frac{d}{dt}\si_1^x)^{t}\longrightarrow -\frac{d}{dt}\si_N^x,\\
(\frac{d}{dt}\si_1^y)^{t}\longrightarrow \frac{d}{dt}\si_N^y,\\
(\frac{d}{dt}\si_1^z)^{t}\longrightarrow -\frac{d}{dt}\si_N^z;
\end{array}
\right\}
\ee
\be
\left.
\begin{array}{l}
(M_2^1)^{t}\longrightarrow M_N^1,\\
(M_2^2)^{t}\longrightarrow -M_N^2,\\
(M_2^3)^{t}\longrightarrow M_N^3;
\end{array}
\right\}
\ee
under the transposition. Here the superscript $t$ denotes the transposion 
of a matrix.\\
 
Note that $(\si^x)^{t}=\si^x, (\si^y)^{t}=-\si^y, (\si^z)^{t}=\si^z$, then from 
the equations (3.6)-(3.11) and the expressions for $M_N^i, M_2^i (i=1,2,3 )$, it 
is straightforward to show that the conditions (3.26) and (3.27) hold. So if we 
get the solutions for $a_{ij}^{(+)}$, using (3.25) the expressions for $a_{ij}^{(-)}$
are also obtained. After a lengthy calculations,  the expressions
for $a_{ij}^{(\pm)l} (i,j=1,2, l=0,1,2,3)$ are as follows,
\ba
a_{11}^{(\pm)0} &=& \frac{4i{\rm  sn}  \eta A_{\pm}}{D} a(u)b(u);\\
a_{11}^{(\pm)1} &=& \frac{i{\rm  sn}  \eta }{D}\{ \pm(B_{\pm}-C_{\pm})c(u)(-2e(u)+f(u)+g(u)) \nonumber \\
                & &  -(B_{\pm}+C_{\pm})(b^2(u)+c^2(u)-f^2(u))\}; \\  
a_{11}^{(\pm)2} &=& \pm\frac{{\rm  sn}  \eta }{D}\{ \pm(B_{\pm}-C_{\pm})(a^2(u)+c^2(u)-e^2(u))\nonumber \\
                & &  +(B_{\pm}+C_{\pm})c(u)(f(u)-g(u))\}; \\
a_{11}^{(\pm)3} &=& G_3-\frac{i{\rm  sn} \eta A_{\pm}}{D}d(u);\\
a_{12}^{(\pm)0} &=& \frac{2i{\rm  sn} \eta }{D} c(u)\{(B_{\pm}-C_{\pm})a(u)+(B_{\pm}+C_{\pm})b(u)\};\\
a_{12}^{(\pm)1} &=& G_1\pm\frac{2i{\rm  sn} \eta A_{\pm}}{D} b(u)(-2e(u)+f(u)+g(u));\\
a_{12}^{(\pm)2} &=&-iG_2\pm \frac{2{\rm  sn} \eta A_{\pm}}{D} a(u)(-f(u)+g(u));\\
a_{12}^{(\pm)3} &=&\mp \frac{i{\rm  sn} \eta }{D}(f(u)+g(u))\{(B_{\pm}-C_{\pm})a(u)
                    +(B_{\pm}+C_{\pm})b(u)\};\\
a_{21}^{(\pm)0} &=&\frac{2i {\rm  sn} \eta }{D}c(u)\{-(B_{\pm}-C_{\pm})a(u)
                    +(B_{\pm}+C_{\pm})b(u)\};\\
a_{21}^{(\pm)1} &=&G_1\mp\frac{2i {\rm  sn} \eta A_{\pm} }{D}b(u)(-2e(u)+f(u)+g(u));\\
a_{21}^{(\pm)2} &=&i G_2\pm\frac{2 {\rm  sn} \eta A_{\pm} }{D}a(u)(-f(u)+g(u));\\
a_{21}^{(\pm)3} &=&\pm\frac{i {\rm  sn} \eta }{D}(f(u)+g(u))\{-(B_{\pm}-C_{\pm})a(u)
                    +(B_{\pm}+C_{\pm})b(u)\};  \\
a_{22}^{(\pm)0} &=&-a_{11}^{(\pm)0} ;\\
a_{22}^{(\pm)1} &=& \frac{i {\rm  sn} \eta }{D}\{ \pm(B_{\pm}-C_{\pm})c(u)(2e(u)-f(u)-g(u)) \nonumber \\
                & &  -(B_{\pm}+C_{\pm})(b^2(u)+c^2(u)-f^2(u))\}; \\
a_{22}^{(\pm)2} &=&\frac{{\rm  sn} \eta }{D}\{ (B_{\pm}-C_{\pm})(a^2(u)+c^2(u)-e^2(u))\nonumber \\
                & &  \mp(B_{\pm}+C_{\pm})c(u)(f(u)-g(u))\}; \\
a_{22}^{(\pm)3} &=&-G_3-\frac{i{\rm  sn} \eta A_{\pm}}{D}d(u)
\ea
where
\ba
a(u) &=& w_2 w_3+w_1w_4,\;\;\; b(u)=w_1 w_3+w_2w_4,\nonumber \\
c(u) &=& w_1 w_2+w_3w_4, \nonumber \\
d(u) &=&(w_2-w_1)^2\{(w_3-w_4)^2-(w_2+w_1)^2\}+ \nonumber\\
     & &   +(w_2+w_1)^2\{(w_3+w_4)^2-(w_2-w_1)^2\},\\
e(u) &=& w_1^2-w_3^2,\;\;\; f(u)=w_2^2-w_3^2,\;\;\;g(u)=w_1^2-w_4^2,\nonumber
\ea
The equations (3.28)-(3.43) combined with the operators $L_n, M_n$  are the Lax pair for the
open $XYZ$ spin-chain. In the trigonometric limit $k\rightarrow 0$, where ${\rm  sn} u\rightarrow \sin u$ 
and taking
$J_x=J_y=1, J_z=\cos \eta, B_{\pm}=C_{\pm}=0$, we recover the result in the case 
of open $XXZ$ model given in the ref. \cite{zhou}(up to a replacement of $\eta$ by $2\eta$).\\

\section{Remarks and discussions}
In this paper, based on the analysis of the symmetries of the reflection 
equations, we present their solutions for the open $XYZ$ model case by case.
We realised that both diagonal and off-diagonal elements of the $K$-matrices
have contributions to the boundary terms of the Hamiltonian, and a specific
scaling factor does not affect the properties of the system. Furthermore, 
we construct the Lax pair for the open $XYZ$ model to show its integrability 
in another way. We believe that our procedure to the solutions of the reflection
equations can be applied to other problems. Recently, the classification of 
six-vertex, eight-vertex solutions of the coloured Yang-Baxter equation has 
been given\cite{sun,wang}, the Yang-Baxter equation with dynamical parameters
(or called the Gervais-Neveu-Felder equation) has been studied\cite{avan}, we 
expect to investegate their corresponding reflection equations on the basis 
of analysing their symmetries.

%\section*{Acknowledgment}

%\newpage


\begin{thebibliography}{99}
\bibitem{fad}L. D. Faddeev, in: {\em Recent Advances in Field Theory and 
Statistical Mechanics, Les Houches XXXIX}, eds. J. B. 
Zuber and R. Stora, North-Holland Publ., 1984, pp 561-608;\\
P. P. Kulish and E. K. Sklyanin, in: {\em Integrable Quantum Field Theories}, 
Lecture Notes in Physics {\bf 151}, Springer-Verlag, 1982, pp 61-119;\\
L. D. Faddeev, {\em How Algebraic Bethe Ansatz Works for Integrable Model},
preprint, hep-th/9605187.

\bibitem{kor}V. E. Korepin, N. M. Bogoliubov and A. G. Izergin, {\em Quantum 
Inverse
Scattering Method and Correlation Functions}, Cambridge University Press,1993.

\bibitem{che}I. Cherednik , {\em Teor. Mat. Fiz.}, {\bf 61}(1984)35.

\bibitem{skl}E. K. Sklyanin,  {\em J. Phys.}, {\bf A 21}(1988)2375.


\bibitem{mez}L. Mezincescu and R. I. Nepomechie,  {\em Int. J. Mod. Phys.}, {\bf A 
6}(1991)5231;
{\bf A 7}(1992)5657.

\bibitem{foe}A. Foerster and M. Karowski, {\em Nucl. Phys.}, {\bf B 408}(1993)512.

\bibitem{gon}A. Gonz\'alez-Ruiz, {\em Nucl. Phys.}, {\bf B 424}(1994)468.

\bibitem{gho}S. Ghoshal and A. Zamolodchikov,  {\em Int. J. Mod. Phys.}, {\bf A 
9}(1994)3841;\\
S. Ghoshal,  {\em Int. J. Mod. Phys.}, {\bf A 9}(1994)4801.

\bibitem{vega1}H J. de Vega and A. Gonz\'alez-Ruiz,  {\em J. Phys.}, {\bf A 
26}(1993)L519.

\bibitem{vega2}H J. de Vega and A. Gonz\'alez-Ruiz, {\em J. Phys.}, {\bf A 
27}(1994)6129.

\bibitem{ina}T. Inami and H. Konno, {\em J. Phys.}, {\bf A 27}(1994)L913.

\bibitem{zhou}H. Q. Zhou, {\em J. Phys.}, {\bf A 29}(1996)L489.

\bibitem{guan}X. W. Guan, M. S. Wang and S. D. Yang, {\em Nucl. Phys.}, {\bf B 485}
(1997)685.

\bibitem{cue}R. Cuerno and A. Gonz\'alez-Ruiz, {\em J. Phys.}, {\bf A 26}
(1993)L605.

\bibitem{sogo}K. Sogo and M. Wadati, {\em Prog. Theor. Phys.}, {\bf 68}(1982)85. 

\bibitem{sun}X. D. Sun, S. K. Wang and K. Wu, {\em J. Math. Phys.}, {\bf 36}
(1995)6043.

\bibitem{wang}S. K. Wang, {\em J. Phys.}, {\bf A 29}(1996)2259.

\bibitem{avan}J. Avan, O. Babelon and E. Billey, {\em Commun. Math. Phys.},
{\bf 178}(1996)281.


\end{thebibliography}
\end{document}